\documentclass[twoside,twocolumn]{article}
\usepackage{jeffe}
\usepackage[nocopyright,toppagenumbers]{proceedings}
\usepackage{smallbib}\bibsize{\footnotesize}
\usepackage{epsfig}
\usepackage{url}

\newtheorem{lemma}{Lemma}[section]
\newtheorem{theorem}[lemma]{Theorem}

\bibalias{Dud}{d-mescs-74}
\bibalias{GJK}{gjk-fpcdb-88}
\bibalias{LC}{lc-eaidc-91}
\bibalias{Lin}{l-efdar-93}
\bibalias{MirV}{m-vclip}
\bibalias{Mir}{m-ibdsrbs-96}
\bibalias{MC}{mc-ibds-95}
\bibalias{DHKS}{dhks-isccd-90}
\bibalias{HS}{hs-parss-95}
\bibalias{Hub}{h-cdiga-95}
\bibalias{Ore}{o-csqpt-90}
\bibalias{Cam}{c-egcmp-97}	
\bibalias{CW}{cw-qcdpv-96}	
\bibalias{BCGMT}{bcgmt-bthrs-96}
\bibalias{GLM}{glm-othsr-96}	
\bibalias{ZF}{zf-berte-95}	

\begin{document}

\title{Separation-Sensitive Collision Detection for Convex Objects}

\urldef{\JEurl}\url{http://www.uiuc.edu/ph/www/jeffe/}
\urldef{\LGurl}\url{http://graphics.stanford.edu/~guibas/}
\urldef{\JSurl}\url{http://www.dcc.unicamp.br/~stolfi/}
\urldef{\LZurl}\url{http://graphics.stanford.edu/~lizhang/}

\author{
	Jeff Erickson\thanks{
		Center for Geometric Computing,
		Department of Computer Science,
		Duke University,
		Durham, NC 27708-0129,
		and Department Computer Science,
		University of Illinois,
		Urbana, IL 61801-2987;
		jeffe@cs.uiuc.edu; \JEurl.
		Portions of this research were done at the
		International Computer Science Institute, Berkeley,
		CA.  Research partially supported by National Science
		Foundation grant DMS-9627683 and by U.S. Army Research
		Office MURI grant DAAH04-96-1-0013.
	}
	\and
	Leonidas J.  Guibas\thanks{
		Computer Science Department,
		Stanford University,
		Stanford, CA 94305;
		guibas@cs.stanford.edu; \LGurl.
                Research partially supported by National Science
                Foundation grant CCR-9623851 and by U.S. Army Research
                Office MURI grant DAAH04-96-1-0007.
	}
	\and
	Jorge Stolfi\thanks{
		Instituto de \Computacao,
		Unversidade Estadual de Campinas,
		Caixa Postal 6065,
		13081-970 Campinas, SP, Brasil;
		stolfi@dcc.unicamp.br; \JSurl.
		Research partially supported by CNPq grant 301016/92-5.
	}
	\and
	Li Zhang\thanks{
		Computer Science Department,
		Stanford University,
		Stanford, CA 94305;
		lizhang@cs.stanford.edu; \LZurl.
	}
}

\date{Submitted to SODA '99\\July 7, 1998}
\date{}

\maketitle
\thispagestyle{empty}
\NotSoLarge	

\begin{abstract}
We develop a class of new kinetic data structures for collision
detection between moving convex polytopes; the performance of these
structures is sensitive to the separation of the polytopes during
their motion.  For two convex polygons in the plane, let $D$ be the
maximum diameter of the polygons, and let $s$ be the minimum distance
between them during their motion.  Our separation certificate changes
$O(\log(D/s))$ times when the relative motion of the two polygons is a
translation along a straight line or convex curve, $O(\sqrt{D/s})$ for
translation along an algebraic trajectory, and $O(D/s)$ for algebraic
rigid motion (translation and rotation).  Each certificate update is
performed in $O(\log(D/s))$ time.  Variants of these data structures
are also shown that exhibit \emph{hysteresis}---after a separation
certificate fails, the new certificate cannot fail again until the
objects have moved by some constant fraction of their current
separation.  We can then bound the number of events by the
combinatorial size of a certain cover of the motion path by balls.
\end{abstract}


\section{Introduction}

Collision detection is an algorithmic problem arising in all areas of
computer science dealing with the simulation of physical objects in
motion.  Examples include motion planning in robotics, virtual reality
animations, computer-aided design and manufacturing, and computer
games.  Often the problem is broken up into two parts, the so-called
\emph{broad phase}, in which we identify the pairs of objects we need
to consider for possible collision, and the \emph{narrow phase} in
which we track the occurrence of collisions between a specific pair of
objects~\cite{Hub}.  (In the spatial database literature, these are
also called the \emph{filtering} and \emph{refinement} phases,
respectively \cite{Ore}.)  For the broad phase, almost all authors use
some kind of simple bounding volumes for the objects themselves, or
for portions of their trajectories in space or space-time, so as to
quickly eliminate from consideration pairs of objects that cannot
possibly collide.  The narrow phase is more specialized, according to
the types of objects being considered.

The simplest objects to consider are convex polytopes (polygons in the
plane, or polyhedra in 3-space), and this case has been extensively
considered in the literature \cite{LC,Mir,MC,GJK,CW}.  More complex
objects are then broken up into convex pieces, which are tested
pairwise.  Algorithmically, the convex polytope intersection problem
is a special case of linear programming; in two and three dimensions
even more efficient techniques have been developed in computational
geometry, that can be applied after a suitable preprocessing of the
two polytopes \cite{DHKS,dk-fdpi-83}.  The methods, however, that have
proven to work best in practice exploit the \emph{temporal coherence}
of the motion to avoid doing an \emph{ab initio} intersection test at
each time step.  Not surprisingly, the collision detection problem is
closely related to the \emph{distance computation} problem.  Since the
distance between two continuously moving polytopes also changes
continuously, many well-known collision detection algorithms, such as
those of Lin and Canny~\cite{Lin,LC}, Mirtich~\cite{Mir,MirV,MC}, and
Gilbert \etal~\cite{GJK} (see also \cite{Cam}), are based upon
tracking the closest pair of features of the polytopes during their
motion (which, of course, implies knowledge of the distance between
the polytopes).  The efficiency of these algorithms is based on the
fact that, in a small time step, the closest pair of features will not
change, or will change to some nearby features on the polytopes.

Though it is hard to imagine how one can do better than tracking the
closest pair of feature when the polytopes are in close proximity,
such tracking seems to be unnecessarily complicated when the polytopes
start moving further from each other.  Indeed most of the above
authors suggest performing first a simple bounding volume (box or
sphere) test on the two polytopes, and only if that fails entering the
closest feature pair tracking mode.  In this paper we consider a
number of general techniques that allow us to perform collision
detection between two moving convex polytopes in a way that is
\emph{sensitive to the separation} between the polytopes.  In order to
properly quantify the separation-sensitivity of our methods, we view
the collision detection problem in the context of \emph{kinetic data
structures} (or \emph{KDSs} for short), introduced in
\cite{bgh-dsmd-97,g-kdssa-98}.

In the kinetic setting we assume that the instantaneous motion laws
for our polytopes are known, though they can be changed at will by
appropriately notifying the KDS.  Our sampling of time is not fixed,
but is determined by the failure of certain conditions, called
\emph{certificates}.  In our case these are \emph{separation
certificates}, which prove that the two polytopes do not intersect.
The failure of a separation certificate need not mean that a collision
has occurred; it can simply mean that that certificate has to be
replaced by one or more others, still proving the non-intersection of
the polytopes.  A good KDS is \emph{compact} if it requires little
space, \emph{responsive} if it can be updated quickly after a
certificate failure, \emph{local} if it adjusts easily to changes in
the motion plans of the objects, and \emph{efficient} if the total
number of events is small.  Our kinetic collision-detection data
structures have all these properties; they maintain only a small
constant number of certificates, and the cost for processing a
certificate failure or a motion plan update is at most polylogarithmic
(in the combinatorial size of the polytopes).

A number of the papers referenced above make the claim that their
algorithms are efficient because ``if the sampling interval is small
enough, then the cost of updating the closest pair of features is
$O(1)$.''  This is a difficult statement to attach a fully rigorous
meaning to---exactly how small the time step must be to guarantee this
condition depends a great deal on both the polytopes and the speed and
complexity of their motion.  In the kinetic setting we can give the
notion of efficiency a more satisfactory theoretical definition, by
focusing on the maximum number of certificate failures we may have to
process for polytopes and motions of a certain complexity, rather than
on the adequacy of any absolute unit of time.

The key contribution of this paper is to develop a class of new
kinetic data structures for collision detection between convex
polytopes, where the efficiency of the structure can be analyzed in
terms of natural attributes of the motion.  Given two moving convex
polygons in the plane, let $\mu=\min\set{n, \sqrt{D/\sigma}}$, where
$n$ is the combinatorial complexity of the polygons, $D$ is their
maximum diameter, and $\sigma$ is their minimum separation during the
entire motion.  In Section~\ref{S:sensitive} we develop a KDS where
the number of events (certificate failures) is $O(\log\mu)$ when the
relative motion of the two polygons is translation along a convex
trajectory (for example, a straight line), $O(\mu)$ for translation
along a algebraic trajectory, and $O(\mu^2)$ for algebraic rigid
motion (translation and rotation).  Thus we see how the nature of the
motion, as well as the proximity of the two polygons, affect the
complexity of the collision detection problem.

In contrast to this, the closest pair of features of two polygons can
change $\Omega(n^2)$ times under an algebraic rigid motion, no matter
what their separation is.  For `intermediate separation' situations,
when the bounding boxes of two objects intersect, but their distance
is still $\Omega(D)$, our methods will perform much better than other
extant methods for collision detection.  The performance of our
methods interpolates smoothly those of the bounding box and closest
pair of features techniques mentioned above, as the separation varies.
In this intermediate distance range our methods are also directly
useful for non-convex objects, as such objects can be bounded by their
convex hulls.

We attain these distance-sensitive bounds by constructing certain
novel outer approximating hierarchies for our polytopes, whose
structure is of independent interest. These hierarchies provide a
series of combinatorially simpler and simpler shells, as we move away
from the polytope. For two polytopes in proximity the hierarchies
locally refine so as to provide a separation certificate.

Hidden in the above `$O$' bounds are factors depending on the
algebraic degree of the motion.  Again, when the polytopes are in
close proximity, it is clear why a `wiggly' motion should be more
costly than a smooth one.  But why should it be so when they are
further away?  This has motivated us to develop structures that
exhibit \emph{hysteresis}---where, after a certificate failure has
occurred, no other certificate failure can happen until the objects
have moved by some constant fraction of their current separation.
Using these structures, we are able to bound the number of events by
the combinatorial size of a certain cover of the motion path by balls
(Section~\ref{S:hysteresis}), somewhat reminiscent of
\cite{mms-qsrs-94}.

We believe that the KDSs shown here are of theoretical and practical
interest.  A basic tool for all our structures are certain outer
approximation hierarchies for convex polytopes and their Minkowski
sums---a topic which we believe to be of independent interest
(Section~\ref{S:hierarchies}).  The distance- and motion-sensitive
bounds we give are novel and, to our knowledge, the first such to be
ever presented.
\enlargethispage{2pt}	

It was surprising to us that even for the simple setting of two moving
convex polygons, there is much that is novel and interesting to say;
in fact, many challenging open questions remain.  Though our
exposition is focussed on the two-dimensional case, we do not
anticipate significant obstacles in extending our results to three
dimensions; we briefly describe some preliminary results in Section
\ref{S:outro}.  We expect that our kinetic structures will lead to
improved practical algorithms for convex shapes, and we have already
started an implementation of our algorithms.  In a companion paper
\cite{beghz-kcdts-98}, we discuss a different set of kinetic collision
techniques applicable to non-convex shapes.

\section{Models of motion}
\label{S:motion}

In our model, each object is a closed rigid convex polygon, whose
motion is described by a moving orthogonal reference frame: a point
$o(t)$ and two orthogonal unit vectors $x(t),y(t)$, whose coordinates
are continuous algebraic functions of~$t$.  (Such moving frames do
exist, and are flexible enough to approximate any motion to any order
and accuracy, for a limited time.  Note however that an algebraic
rotation is necessarily of non-uniform angular velocity, and can cover
only $O(d)$ full turns, where $d$ is the degree of the entries.)  Each
vertex of the object is assumed to have constant algebraic coordinates
$(X,Y)$ relative to the frame $(o,x,y)$; so that its position at
time~$t$ is $o(t) + X x(t) + Y y(t)$, also an algebraic function
of~$t$.

As we shall see, the certificates used in each of our KDSs have the
general form $F(t) = G(p_1(t), \dots, p_n(t))$. Here $p_1(t), \dots,
p_m(t)$ are the positions of $m$ vertices ($m$ will be a small
constant), possibly on different objects; and $G$ is some algebraic
function.  Then $F$ itself is an algebraic function, which means we
can compute the next time~$t$ when $F(t)=0$, exactly, and compare any
two times, at finite cost, within an appropriate arithmetic model.
Moreover, the number of zeros of $F$ in any finite interval is bounded
by its algebraic degree, which is the degree of $G$ times the maximum
algebraic degree~$d$ among the motion coordinates.  So, for, example,
the same vertex triplet cannot become collinear more than $2d$ times.

In conclusion, if the complexity (algebraic degree and coefficient
size) of the motions is bounded, each certificate can fail at most
$O(1)$ times during a single motion, and the cost of computing and
comparing the failure times is also $O(1)$.  We will use these last
two facts, and \emph{only} these two facts, throughout our analyses.
Our results do not require any other combinatorial properties of
algebraic motion---for example, that an algebraic path can be
decomposed into a constant number of convex sub-paths, as in the
companion paper \cite{beghz-kcdts-98}---and therefore apply to a wider
range of \emph{pseudo-algebraic} motions~\cite{bgh-dsmd-97,g-kdssa-98}.

\section{Polygon Approximation Hierarchies}
\label{S:hierarchies}

Our collision detection algorithms are based on outer approximation
hierarchies for the convex polygons involved, or for their Minkowski
sum.  All these hierarchies tile the space exterior to the polygon so
as to simplify the combinatorial structure of the polygon as one goes
away from the polygon.  A well-known example of such a hierarchy in
computational geometry is the Dobkin-Kirkpatrick hierarchy
\cite{dk-fdpi-83}.  This hierarchy is not by itself adequate for our
purposes, however, as it is not sensitive to the distance away from
the polygon; any approximation in the hierarchy can have vertices
arbitrarily far from the original polygon.  We will retain a key
property of this hierarchy, namely the fact that that each `coarsening
step' is performed by removing an existing edge of the current
approximation and extending outwards its two neighboring edges till
they meet.  However, in order to guarantee that we get the
distance-sensitivity that we require, we will enrich the set of
`available edges' by introducing an additional set of degenerate edges
around the polygon, initially all of zero length.  In general, the
number of these degenerate edges will be proportional to the number of
original edges in the polygon.

\begin{figure}
\centerline{\epsfig{file=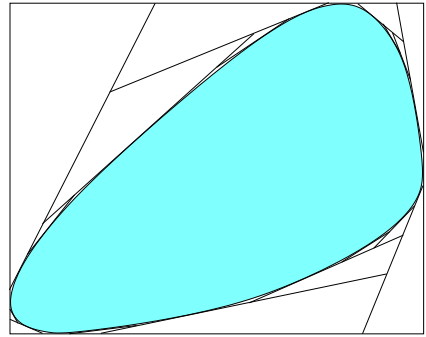,height=1.75in}}
\caption{A boomerang hierarchy.}
\label{boom}
\end{figure}

Though one normally thinks of these approximating hierarchies as being
constructed from the boundary of the polygon towards the outside, it
is actually advantageous to visualize this process in reverse.  Let
$P$ be our convex polygon of $n$~edges.  We start from the outermost
approximation $P_0$, which for our purposes is always a (not
necessarily axis-aligned) bounding rectangle of~$P$.  The space
between $P$ and $P_0$ comprises up to four non-convex polygons, each
consisting of a concave chain and two additional edges.  Following
Hershberger and Suri \cite{HS}, we call these polygons
\emph{boomerangs}.  The last coarsening step, if undone, corresponds
to `cutting off a corner' of~$P_0$ through the reintroduction of
another (possibly degenerate) edge of~$P$.  This step splits one of
the top-level boomerangs into a triangle and two smaller boomerangs.
This process of cutting off corners is continued recursively, until
all edges of~$P$ have been reintroduced.  See Figure~\ref{boom}.
Structurally, the hierarchy consists of four binary trees, where each
internal node corresponds to a triangle and each subtree to a
boomerang.  In our hierarchies, each of these trees will have depth
$O(\log n)$.  A variety of outer approximations for $P$ can be defined
by removing a subtree of triangles rooted at each of the top-level
boomerangs.

Before we discuss various choices of degenerate edges, let us develop
some notation and terminology.  The \emph{apex} of a boomerang is the
unique vertex not on the boomerang's concave chain.  The \emph{height}
of a boomerang is the distance from its apex to its concave chain;
this is the maximum distance from any point in the boomerang to the
chain.  The \emph{level} of either a boomerang or a triangle is its
depth from the root in the appropriate binary tree; there are at most
$4\cdot 2^i$ boomerangs at level $i$.  Finally, let $P_i$ denote the
$i$-th \emph{envelope} of~$P$, defined as the union of~$P$ and all
level-$i$ boomerangs; $P_i$ is itself a convex polygon
surrounding~$P$.

Let $D$ denote the diameter of~$P$.  The following lemma implies that
the envelopes in any boomerang hierarchy of~$P$ are reasonably close
to~$P$.  We omit the easy proof from this abstract.

\begin{lemma}\label{L:fewtall}
In any single level of any boomerang hierarchy, there are
$O\big(\sqrt{D/s}\big)$ boomerangs of height at least $s$.
\end{lemma}

A line $\ell$ that does not intersect $P$ can intersect at most one of
the four top-level boomerangs in any boomerang hierarchy of~$P$.
Moreover, if $\ell$ intersects a boomerang, then it intersects at most
one of its two sub-boomerangs.  In fact, these two observations hold
for any convex curve (bending away from $P$).  These simple
observations establish the following useful lemma.

\begin{lemma}\label{L:path}
Any convex curve disjoint from $P$ intersects at most one triangle in
each level of any boomerang hierarchy of~$P$.
\end{lemma}

We observe that the triangles in a boomerang hierarchy for $P$ tile
the space between $P_0$ and $P$.  (It is not hard to extend this
tiling to be the complement of~$P$ in the plane by allowing a few
infinite triangles.  This is almost identical to the construction of a
binary space partition tree \cite{fkn-vsgpt-80}.)  For a point $x$
outside $P$, the triangle in this tiling that contains $x$ provides us
with useful information about the position of $x$ with respect to $P$:
the base of the triangle is a side of~$P$ separating $x$ from $P$,
while the height of the associated boomerang is an upper bound on the
distance from $x$ to $P$.

\subsection{The Compass Hierarchy}

For the \emph{compass hierarchy} we introduce $O(n)$ zero-length edges
into $P$, whose outer normals form a regular recursive lattice on the
unit circle, in the standard compass directions (E, N, W, S, NE, NW,
SW, SE, etc.).  In the top $\ceil{\log_2 n}$ levels of the compass
hierarchy, each boomerang is subdivided into two smaller boomerangs
and an \emph{isosceles} triangle.  In the remaining levels, if any,
each boomerang is subdivided by a line through its median edge, as in
a standard Dobkin-Kirkpatrick hierarchy.  The resulting hierarchy has
at most $2\ceil{\log_2 n}$ levels.

\begin{lemma}\label{L:compass}
{\bf (a)} A level-$i$ boomerang in the compass hierarchy of $P$ has
height $O(D/2^i)$.

{\bf (b)} The compass hierarchy of $P$ contains $O\big(\sqrt{D/s}\,
\log(D/s)\big)$ boomerangs with height at least~$s$.

{\bf (c)} Any convex curve at distance $s$ from $P$ intersects
$O(\log(D/s))$ triangles in the compass hierarchy of $P$.

{\bf (d)} If two polygons $P$ and $Q$ are distance $s$ apart, their
compass hierarchies contain disjoint approximations $\widetilde{P}$
and $\widetilde{Q}$, each with $O(\log(D/s))$ edges.
\end{lemma}

Klosowski \etal~\cite{khmsz-ecdubv-98} define the ``$k$-DOP'' or
\emph{discrete orientation polytope} of an object to be the bounding
polytope whose facets are normal to a fixed set of $k$ `compass'
directions.  Klosowski \etal\ construct a hierarchy of bounding
volumes for any object by computing the object's $k$-DOP, decomposing
the object into a constant number of pieces, and recursively
constructing a hierarchy for each piece, using the same value of $k$
at all levels.  (See \cite{BCGMT,GLM,ZF} for similar bounding volume
hierarchies.)  In contrast, the compass hierarchy consists of a nested
sequence of $k$-DOPs with $k=4,8,16,\dots$, all for the same object.

\subsection{The Dudley Hierarchy}

\begin{figure}
\centerline{\epsfig{file=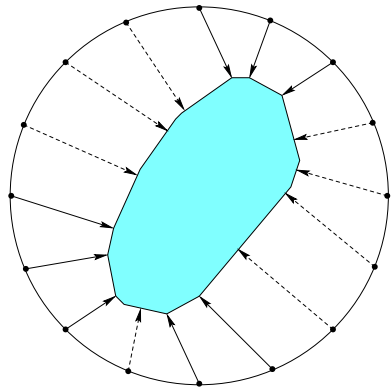,height=2in}}
\caption{The Dudley construction.  Solid vectors introduce degenerate
edges; dashed vectors introduce degenerate vertices.}
\label{dudley2}
\end{figure}

Our second boomerang hierarchy is based on a result of
Dudley~\cite{Dud} on approximating convex bodies in arbitrary
dimensions by polytopes with few facets; hence, we call it the
\emph{Dudley hierarchy}.  Let $S$ be a set of $n$ regularly spaced
points on a circle of radius $2D$, centered inside $P$.  For each
point $x \in S$, let $n(x)$ be its nearest neighbor on $P$.  If $n(x)$
is a vertex of~$P$, we introduce a zero-length edge at $n(x)$ whose
outer normal vector is $n(x)-x$.  Otherwise, we introduce a new
degenerate vertex at $n(x)$ whose external angle is zero.  See
Figure~\ref{dudley2}.  We then create a boomerang hierarchy, starting
with the bounding box $P_0$, by recursively subdividing each boomerang
by a line through its median (possibly degenerate) edge (and possibly
through other collinear edges).  The resulting \emph{Dudley hierarchy}
has depth at most $\ceil{\log_2(2n)}$.

\begin{lemma}\label{L:dudley}
{\bf (a)} A level-$i$ boomerang in the Dudley hierarchy of $P$ has
height $O(D/4^i)$~\cite{Dud}.

{\bf (b)} The Dudley hierarchy of $P$ contains $O\big(\sqrt{D/s}\big)$
boomerangs with height at least $s$.

{\bf (c)} Any convex curve at distance $s$ from $P$ intersects
$O(\log(D/s))$ triangles in the Dudley hierarchy of $P$.

{\bf (d)} If two polygons $P$ and $Q$ are distance $s$ apart, their
Dudley hierarchies contain disjoint approximations $\widetilde{P}$ and
$\widetilde{Q}$, each with $O(\log(D/s))$ edges.
\end{lemma}

%
\section{Mixed Hierarchies}
\label{S:mixed}

The hierarchies introduced so far are tilings of the free space around
a single convex polygon.  Since we are interested in systems with two
(or more) moving convex polygons, we would like to extend some of
these notions to hierarchical tilings of the free part of the
configuration space generated by the motion of two polygons.  Our
general plan is \emph{to associate} (explicitly or implicitly) \emph{a
particular separation proof with each tile of such a tiling}.  As the
two polygons move and their configuration crosses out of its current
tile, some certificates will fail and a new separation certificate
will have to be generated.

Let $P$ and $Q$ be our two moving polygons.  If $P$ and $Q$ are only
translating with respect to each other, then the configuration space
remains two-dimensional.  It is well-known that in this case the free
space is the complement of the convex polygon $P \oplus (-Q)$, the
Minkowski sum of $P$ with the negative of $Q$.  Since again our
two-dimensional configuration space is the exterior of a convex
polygon, then all the hierarchies presented above can be directly
used, if we are willing to construct this polygon.  Such a direct
approach, however, is less attractive when rotation is allowed, as
then the configuration space can have quadratic combinatorial
complexity.  Also, for applications where we have multiple moving
polygons (though we do not discuss techniques for this many-body
problem in this paper), we want to `recycle' as much as possible
pieces of the hierarchies built for each of the polygons, rather than
having to build a separate hierarchy for each pair.

We describe below one such hierarchy for tiling the complement of the
Minkowski sum of two convex polygons into triangles and
parallelograms, which we call the \emph{mixed hierarchy}.  (Without
loss of generality, we focus on describing this hierarchy for $P
\oplus Q$, as opposed to $P \oplus (-Q)$.)  It is motivated by the
theory of mixed volumes of convex bodies \cite{s-cbbmt-93}, and it has
the advantage that it changes in a very regular way as the polygons
rotate.  This makes it possible to maintain a non-intersection
certificate for two moving convex polygons by separating the motion
into a \emph{translational} part, corresponding to a point moving
around in the plane, and a \emph{rotational} part, corresponding to a
deformation of the triangle or parallelogram containing the point.  At
certain discrete events this tile and a neighboring tile are deleted
and replaced by two other tiles covering the same area, much like a
Delaunay flip.  Thus, we can maintain a separation proof by
maintaining one tile, or a small working set of tiles, around the
current configuration point.

\begin{figure}
\centerline{\epsfig{file=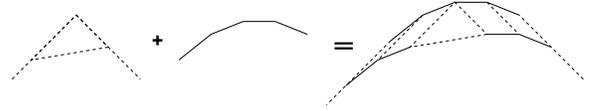,width=3in}}
\caption{Inserting an edge into the mixed hierarchy.}
\label{mixed}
\end{figure}

Given two boomerang hierarchies for $P$ and~$Q$, the mixed hierarchy
is defined by starting from $P_0 \oplus Q_0$ and then interleaving the
`corner cutting' operations leading to $P$ and~$Q$.  Different mixed
hierarchies may be obtained for different interleavings of these
corner cutting operations.  The main difference from a standard
boomerang hierarchy arises because the corner of $P$ or~$Q$ being cut
next may no longer be a corner at all in the Minkowski sum hierarchy:
between two consecutive sides of~$P$ we can have many sides of~$Q$,
and vice versa.  To be concrete, say the next cut is to add a side~$e$
of~$P$, but its neighboring sides in the $P$ hierarchy are now
separated by a chain of $Q$ edges in the mixed hierarchy built so far.
Note that these neighboring sides must already be present in the mixed
hierarchy, as the joint corner cutting sequence is consistent with
that for $P$.  To insert $e$ into the mixed hierarchy, we fist find
the place where $e$ fits, according to slope, in the chain of
$Q$-edges that replaced its corner.  We partition that $Q$-chain at
that point, and then translate the two subchains inwards, as
illustrated in Figure~\ref{mixed}.  The two chains come to rest
when they encounter the points where $e$~meets its two neighboring
edges in $P$.  Thus this process adds to the mixed hierarchy exactly
the same triangle that it added in the $P$ hierarchy, as well as
several parallelograms based on the edges of the $Q$ chain.  After all
the corners have been cut, the space between $P_0 \oplus Q_0$ and $P
\oplus Q$ is tiled with one copy of each of the original triangles
used in the $P$ and $Q$ hierarchies, as well as several `mixed'
parallelograms.

\begin{figure}
\centerline{\epsfig{file=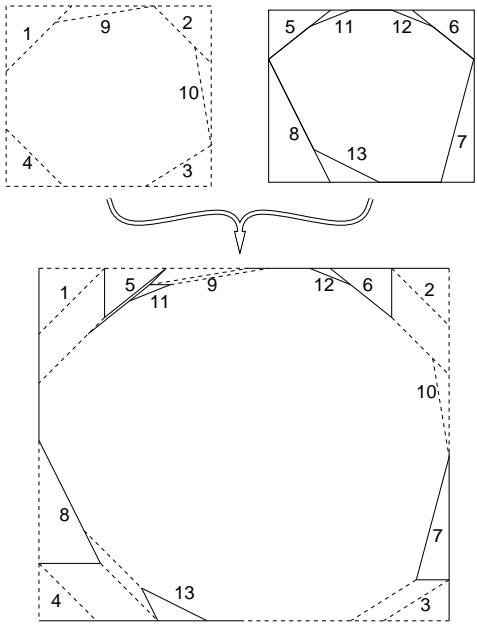,width=2.5in}}
\caption{A mixed hierarchy for two convex polygons.}
\label{full-mixed}
\end{figure}

Figure~\ref{full-mixed} shows a full mixed hierarchy for two small
convex polygons.  The corner cutting order is indicated by the numbers
next to the edges of the polygons.

\begin{figure}
\centerline{\epsfig{file=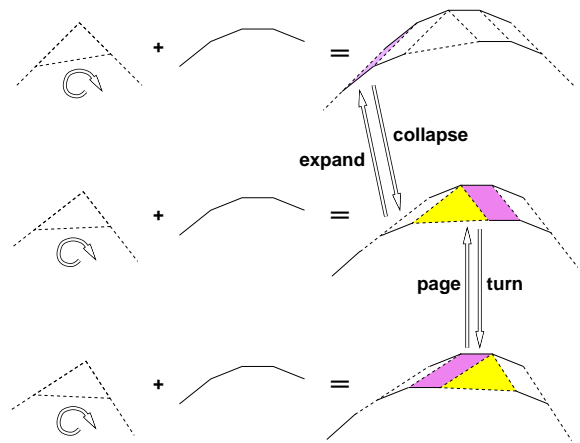,width=3in}}
\caption{Events in the mixed hierarchy under rotation.}
\label{mixed-events}
\end{figure}

As we mentioned earlier, a nice aspect of the mixed hierarchy is its
behavior under polygon rotation.  The rotation changes the
interleaving of the edges of $P$ and $Q$ around their Minkowski sum.
For example, an edge of $P$ can pass in slope ordering one of the
edges of the $Q$-chain that it splits.  The primary effect of this
change on the mixed hierarchy is a simple change, which we call a
\emph{a page turn}, akin to a Delaunay flip: a triangle and a
parallelogram exchange positions.  Additionally, at some finer level
of the hierarchy, one parallelogram may either \emph{collapse} to a
line segment and be destroyed, or a new parallelogram may be created
and \emph{expand} from a line segment.  See Figure~\ref{mixed-events}.

\begin{lemma}\label{L:mixed-size}
If the corner cutting operations for polygons $P$ and $Q$ of size $n$
and $m$, where $m\le n$, are interleaved according the level of the
boomerang destroyed at each step (in the $P$ or $Q$ hierarchy
respectively), then the size of the mixed hierarchy is $O((m+n) \log
m)$.
\end{lemma}

\begin{lemma}\label{L:mixed-rotation}
If $P$ is stationary and $Q$ makes a single full rotation, then the
number of page turns that will happen is $O(mn \log m)$.
\end{lemma}

\begin{lemma}\label{L:mixed-onecell}
Given a cell $c$ of the mixed hierarchy, a point $p\in\partial c$, and
some auxiliary structures of linear size for $P$ and $Q$, the
neighboring cell of $c$ containing $p$ can be computed in $O(\log n)$
time.  The same applies for the cells covering $c$, when cell $c$ is
destroyed during a page turn event.
\end{lemma}

If the boomerang hierarchies of $P$ and $Q$ satisfy distance
properties such as those in Section~\ref{S:hierarchies}, then similar
results will hold for the tiles of the mixed hierarchy.  For example,
if we mix two compass or Dudley hierarchies, we can reduce the time
bound in Lemma \ref{L:mixed-onecell} from $O(\log n)$ to $O(\log
(D/s))$, where $s$ is the distance from $p$ to $P\oplus Q$.

In general, however, the mixed hierarchy does \emph{not} satisfy the
equivalent of Lemma~\ref{L:path}.  Although any line disjoint from
$P\oplus Q$ hits only one \emph{triangle} per level, it may hit a
large number of that triangle's neighboring \emph{parallelograms}.
Since the compass hierarchy uses the same evenly-spaced cutting
directions for every polygon, at most one parallelogram appears next
to any triangle in the mixed compass hierarchy, so this problem is
avoided.  Unfortunately, we have no similar guarantee for the Dudley
hierarchy, so our bound in that case is much weaker.

\begin{lemma}\label{L:mixed2}
{\bf (a)} The mixed compass hierarchy and the mixed Dudley hierarchy
of ${P\oplus Q}$ have $O\big(\sqrt{D/s}\,\log(D/s)\big)$ and
$O\big(\sqrt{D/s}\big)$ vertices with height at least~$s$,
respectively.

{\bf (b)} Any convex curve at distance $s$ from $P\oplus Q$ intersects
$O(\log(D/s))$ and $O\big(\sqrt{D/s}\big)$ triangles in the mixed
compass hierarchy and mixed Dudley hierarchy of $P\oplus Q$,
respectively.
\end{lemma}

\section{Separation-Sensitive Data Structures}
\label{S:sensitive}

Several authors have proposed algorithms to maintain the closest pair
of features between two convex polygons \cite{GJK,LC,Mir}; these
algorithms can easily be transformed into a kinetic data structure
with constant update time.  There are at least two alternative
approaches that lead to the same performance.  We could maintain an
\emph{inner common tangent} between $P$ and $Q$, \ie, a line that
touches the boundaries of both objects, but separates their interiors.
Alternately, we could maintain a \emph{separating edge} $e$ of one
polygon, along with the vertex $v$ of the other polygon closest to the
line through $e$.  Unfortunately, for all three approaches, the
worst-case number of kinetic events is quite high---$\Theta(n)$ if the
polygons are only translating, and $\Theta(n^2)$ if they are also
allowed to rotate.  Moreover, these lower bounds can be achieved while
the polygons are arbitrarily far apart.

In this section, we describe several new kinetic data structures that
maintain a separation certificate between two moving convex polygons,
where the cost and number of events depends on the distance between
the polygons.  Our data structures are loosely based on the algorithm
of Dobkin \etal~\cite{DHKS} for detecting intersections between
preprocessed convex polygons or polyhedra.

Let us first establish some notation.  $P$ and $Q$ are convex $n$-gons
with diameter at most $D$.  We let $s$ denote the current separation
(geometric distance) between $P$ and $Q$ at a given time, and $\sigma$
the minimum of $s$ over the entire history of the motion.  Finally, we
let $\mu = \min\set{n, \sqrt{D/\sigma}}$.

\subsection{One Point, One Polygon}

We illustrate our basic approach by considering the special case where
$P$ consists of a single point~$p$.  We construct either a compass or
Dudley hierarchy $Q_0, Q_1, Q_2, \dots$ for $Q$.  Each triangle at
level $i$ in this hierarchy has an \emph{inner} edge, which is an edge
of $Q_{i+1}$, and two \emph{outer} edges, which are subsets of edges
of $Q_i$.

\begin{figure}
\centerline{\footnotesize
\begin{tabular}{cc}
        \epsfig{file=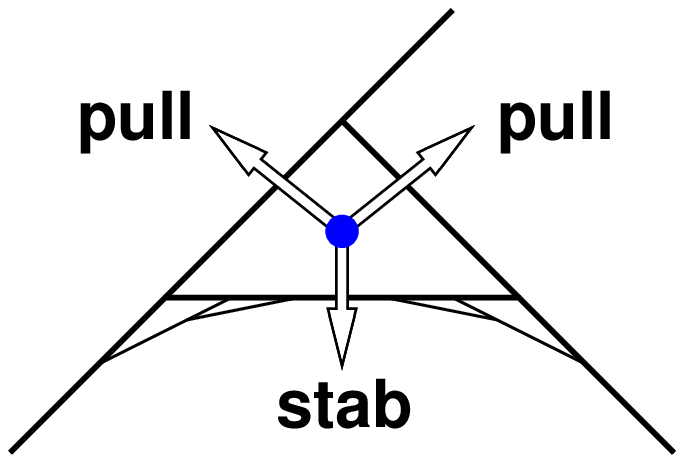,height=0.75in} &
        \epsfig{file=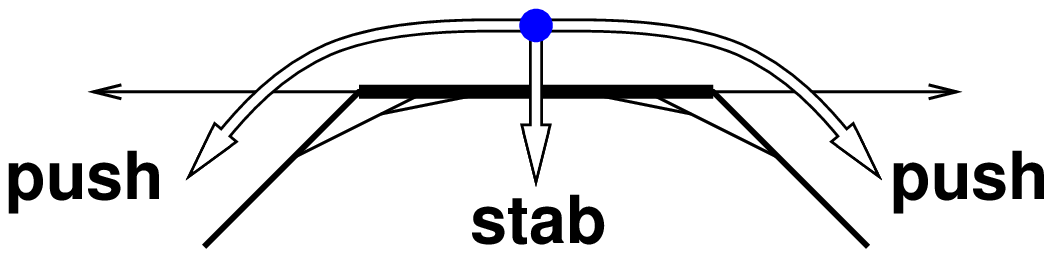,height=0.75in} \\
	(a) & (b)
\end{tabular}}
\caption{Events for a point moving around a polygon.  (a)~Maintaining
the active triangle. (b)~Maintaining a separating edge.}
\label{F:pp-events}
\end{figure}

At any moment, we maintain the \emph{active} triangle $\triangle$
containing the point $p$.  There are two types of certificate
failures; see Figure~\ref{F:pp-events}(a).  If $p$ crosses one of the
outer edges of $\triangle$, we can identify the triangle it enters in
constant time.  On the other hand, if $p$ crosses the inner edge of
$\triangle$, it could either collide with $Q$ or pass into a triangle
at some deeper level of the hierarchy.  We can check for actual
collision in $O(1)$ time be seeing if $p$ lies on the line segment
$e\cap Q$.  Otherwise, we search one level at a time for the new
active triangle $\triangle'$.  If $\triangle$ is at level $i$ and
$\triangle'$ is at level $j$, we find $\triangle'$ in $O(j-i) = O(\log
n)$ steps.  Alternately, if we use binary search, we can find
$\triangle'$ in time $O(\log(\log n-i))$.

If the point is moving along a convex path (curved away from $Q$),
then by Lemma \ref{L:path} at most one triangle in each level is ever
active.  For the Dudley and compass hierarchies, the level of the
deepest active triangle, and thus the number of active triangles, is
$O(\log\mu)$.  We easily observe that the total time spent updating
the active triangle is also only $O(\log\mu)$.

Now suppose the point is moving along some other algebraic path.  For
the compass hierarchy, the number of events is $O(\mu\log\mu)$, and
for the Dudley hierarchy, the number of events is $O(\mu)$.  Both
upper bounds follow directly from the number of triangles that can
contain a point at distance $\sigma$ or higher from $Q$ (Lemmas
\ref{L:compass}(b) and \ref{L:dudley}(b)).  As in the case of convex
motion, these are also upper bounds on the total update time.

\begin{theorem}
\label{Th:ptpoly}
As $p$ moves algebraically about $Q$, we can maintain a separation
certificate maintained in $O(\min\set{\log\log n, \log(D/s)})$ time
per event, using $O(n)$ space and preprocessing time.  Both the number
of events and the total update time is $O(\log\mu)$ if $p$ moves along
a convex curve and $O(\mu)$ otherwise.
\end{theorem}

Despite its good performance, the KDS just described is somewhat
wasteful.  Only the inner edge of the active triangle separates the
moving point from the polygon, so why worry about the other two edges?
This observation motivates the following \emph{lazy} variant of our
KDS, which has exactly the same performance bounds as
Theorem~\ref{Th:ptpoly} in the worst case, but is likely to be more
efficient in practice.

Instead of a triangle, we maintain a single separating edge $e_i$ of
some envelope $Q_i$.  We update the separating edge only when $p$
passes through the line containing $e_i$.  Again, there are two types
of events: if $p$ hits the edge $e_i$, we have a \emph{stab} event,
and otherwise, we have a \emph{push} event.  See Figure
\ref{F:pp-events}(b).  Stab events are handled exactly as in the
previous structure: first check for a real collision, and if no
collision has occurred, find a new separating edge at a deeper level
in the hierarchy.  After a push event, one of the edges of $Q_i$
adjacent to $e_i$, say $e'_i$, is now a separating edge.  However, the
structure of the hierarchy ensures that $e'_i$ is actually a subset of
an edge $e_j$ of some coarser envelope $Q_j$; we take $e_j$ to be the
new separating edge.

\subsection{Two Translating Polygons}
\label{ss:translation}

Now consider the case of two convex polygons $P$ and $Q$ which are
translating along algebraic paths.  It suffices to consider the case
where $Q$ is fixed and only $P$ moves.  Detecting collisions between
$P$ and $Q$ is equivalent to detecting collisions between a single
moving point and the static Minkowski sum $P\oplus(-Q)$, so the bounds
in Theorem \ref{Th:ptpoly} immediately applies to the case of two
translating polygons.

This structure is unsatisfactory, however, since it requires us to
construct a decomposition of the Minkowski sum $P\oplus(-Q)$.  If we
want to collisions among several convex $n$-gons using this approach,
we need $O(n)$ space for every active \emph{pair} of polygons.  In
this section, we describe a modification of our previous data
structures that use a separate hierarchy for each polygon, so that we
only need $O(n)$ space for each \emph{polygon}, plus constant space
for every active pair.

Our first approach is to maintain the \emph{active} cell $c$ of the
mixed hierarchy of $P\oplus(-Q)$ containing the current configuration
point.  Whenever the configuration point leaves $c$, we can compute
the new active cell~$c'$ in $O(\log\mu)$ time (Lemma
\ref{L:mixed-onecell}). We emphasize that it is not necessary to
construct the entire mixed hierarchy explicitly, but only separate
hierarchies for the two polygons.  If we build the mixed hierarchy out
of the Dudley hierarchies of $P$ and $Q$, convex translation can now
cause $O(\mu)$ events.  If we use compass hierarchies instead, convex
translation causes only $O(\log\mu)$ events, but the event bound for
more general translations goes up to $O(\mu\log\mu)$.  As in the
one-point case, the event bounds also bound the total update time.

\begin{theorem}
\label{Th:transpoly}
As $P$ and $Q$ translate algebraically, we can maintain a separation
certificate maintained in $O(\log(\min\set{n, D/s})$ time per event,
using $O(n)$ space and preprocessing time per polygon plus $O(1)$
extra space for the pair.  Both the number of events and the total
update time is $O(\log\mu)$ if $P$ moves along a convex curve
relative to $Q$ and $O(\mu)$ otherwise.
\end{theorem}

We can also achieve these bounds with a suitable modification of our
lazy KDS.  Let $P_0, P_1, \dots$ be a boomerang hierarchy for $P$, and
$Q_0, Q_1, \dots$ a boomerang hierarchy for $Q$.  As the polygons
move, we maintain a separation certificate $(e_i,v_i)$, where $e_i$ is
an edge of $Q_i$ and $v_i$ is the closest vertex of $P_i$ to
$\Line{e_i}$ (the line through $e_i$), or vice versa.  We always use
features of the same level in both hierarchies.  Given a valid
separation certificate $(e_i,v_i)$, we can compute either a finer
certificate $(e_{i+1},v_{i+1})$ or a coarser certificate
$(e_{i-1},v_{i-1})$ in constant time, if one exists, by checking local
neighborhoods of $e_i$ and $v_i$.

\begin{figure}
\centerline{\footnotesize
\begin{tabular}{c}
        \epsfig{file=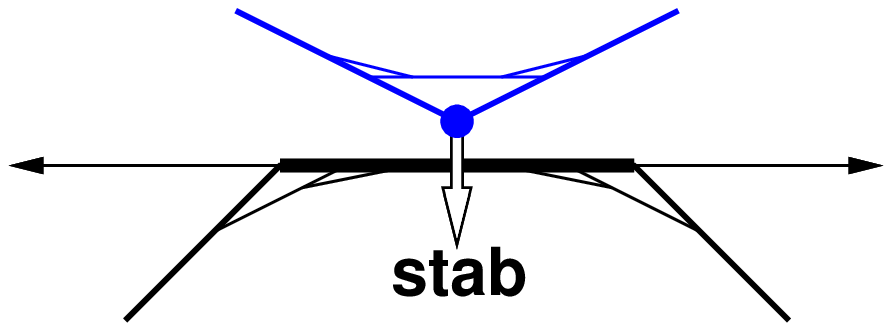,height=0.55in} \quad
        \epsfig{file=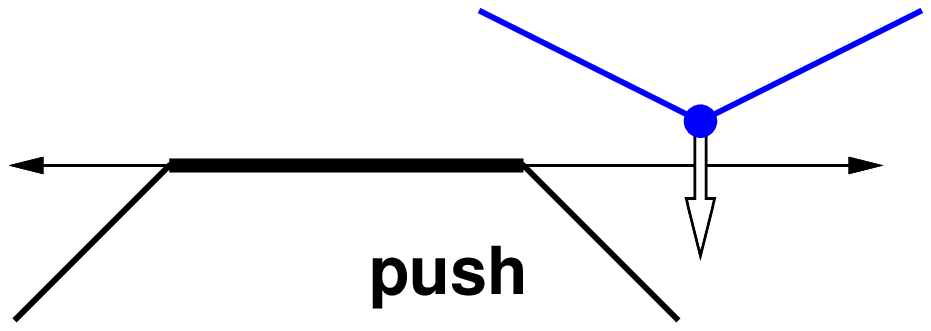,height=0.55in} \\
	(a) \\[2ex]
        \epsfig{file=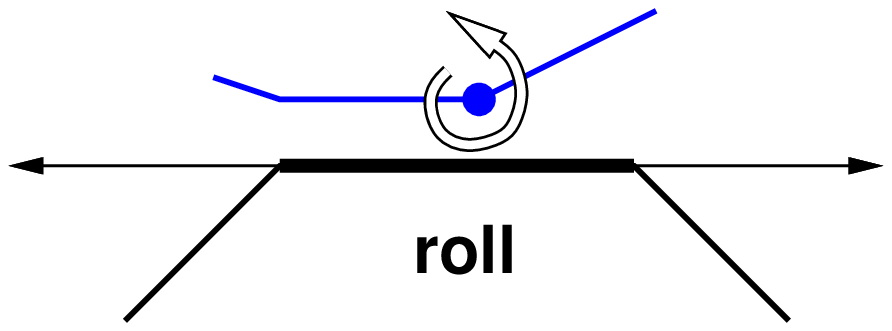,height=0.55in} \\
	(b)
\end{tabular}}
\caption{Events for two polygons undergoing (a)~translation and
(b)~rotation.}
\label{F:2p-events}
\end{figure}

The separation certificate expires when $v_i$ crosses $\Line{e_i}$.
As before, there are two types of events; see Figure
\ref{F:2p-events}(a).  (In the following description, we will assume
without loss of generality that $v_i\in P_i$.)

To handle a \emph{stab} event, where $v_i$ hits $e_i$, we refine both
envelopes one level at a time until they are disjoint.  (Since a
single refinement may introduce a zero-length edge at $v_i$, we may
have to refine by more than one level.)  If $P_j$ and $Q_j$ are the
coarsest disjoint envelopes, then either the edge of $Q_j$ containing
$e_i$ or an edge of $P_j$ containing $v_i$ is a new separating edge.
If $v_i$ is actually a vertex of $P$ and it hits the edge of $Q$
containing $e_i$, we report a collision.

After a \emph{push} event, where $v_i$ does not hit the edge $e_i$,
either an edge of $Q_i$ adjacent to $e_i$ or an edge of $P_i$ adjacent
to $v_i$ is a separating edge between $P_i$ and $Q_i$ (or possibly
both), depending on which has the higher slope.  After updating the
separation certificate, we coarsen the envelopes one level at a time,
computing a new separation certificate at each new level, until the
next coarser envelopes intersect.

Since we can refine or coarsen by one level in constant time, the
total cost of either event is $O(\log\mu)$.  The number of events is
the same as for the mixed-hierarchy structure.

\subsection{Rigid Motion}

Now suppose $P$ and $Q$ are also allowed to rotate.  As we mentioned
earlier, the mixed hierarchy changes as the polygons rotate.  If the
active cell in the mixed hierarchy disappears due to a page turn, we
can construct the new active cell in $O(\log\mu)$ time.  Since we only
maintain the active cell, page turns elsewhere in the mixed hierarchy
cost us nothing.  We do not have to worry about collapse events, since
the configuration point will be 'squeezed out' of the cell before it
finishes collapsing.

Our lazy two-hierarchy data structure can now encounter an additional
event, called a \emph{roll}, when one of the edges adjacent to $v_i$
becomes parallel to $e_i$; see Figure~\ref{F:2p-events}(b).  To handle
a roll event, we keep $e_i$ as the separating edge, but now its
nearest neighbor in the other envelope is a vertex adjacent to $v_i$.
After we update the separation certificate, we then coarsen the
envelopes as much as possible in $O(\log\mu)$ time, just as for a push
event.

For both approaches, we have the following bounds, if we use the
Dudley hierarchy; the event bound for the compass hierarchy is
slightly weaker.

\begin{theorem}
\label{Th:rigidpoly}
As $P$ and $Q$ undergo algebraic rigid motion, we can maintain a
separation certificate in $O(\log(\min\set{n, D/s})$ time per event,
using $O(n)$ space and preprocessing time per polygon, plus $O(1)$
space for the pair.  Both the number of events and the total update
time are $O(\mu^2)$.
\end{theorem}


\subsection{Summary}

Table \ref{T:results} summarizes the event bounds for our kinetic data
structures for various types of objects, classes of motion, and
boomerang hierarchies.

\begin{table}
\centerline{\small
\begin{tabular}{c|cc}
Motion	& Compass	& Dudley	\\
\hline
\hline
\multicolumn{3}{c}{\textbf{One point, one polygon}}
\\
\hline
	 convex		& $O(\log\mu)$	& $O(\log \mu)$ \\
	 general	& $O(\mu\log\mu)$ & $O(\mu)$	\\
\hline
\hline
\multicolumn{3}{c}{\textbf{Two polygons}}
\\
\hline
	 convex translation
			& $O(\log\mu)$	& $O(\mu)$	\\
	 general translation
			& $O(\mu\log\mu)$ & $O(\mu)$	\\
	 rigid motion	& $O(\mu^2\log^2\mu)$	& $O(\mu^2)$	\\
\end{tabular}
}
\caption{The number of events and total update times of our KDSs, for
different types of objects, motions, and boomerang hierarchies.  Here,
$\mu=\min\set{n,\sqrt{D/\sigma}}$, where $D$ is the objects' maximum
diameter and $\sigma$ is their minimum separation.}
\label{T:results}
\end{table}

\section{Inflation, Hysteresis, and Path-Sensitivity}
\label{S:hysteresis}

By further modifying the lazy variants of the kinetic data structures
described in the previous section, we can obtain data structures that
exhibit \emph{hysteresis}: after any event, the configuration must
change by a certain amount before the next event occurs.  Hysteresis
allows us to derive upper bounds on the number of events based on
geometric properties of the path that, unlike our earlier results, do
not depend on smoothness or algebraicity.

For any convex polygon $P$ and real number $\e>0$, the (outer)
\emph{offset polygon} $P[\e]$ is a convex polygon obtained by moving
each edge of $P$ outwards by a distance of $\e$ and moving the
vertices along their angle bisectors.  That is, each edge of $P[\e]$
is parallel to an edge of $P$ and vice versa, and the distance between
their two lines is~$\e$.  For any point $p$ on $P[\e]$, we have $\e
\le d(p,P) \le \e/\sin(\theta/2)$, where $\theta$ is the minimum
internal angle at any vertex of $P$.  Zero-length edges in $P$ induce
positive-length edges in $P[\e]$.

\begin{figure}
\centerline{\epsfig{file=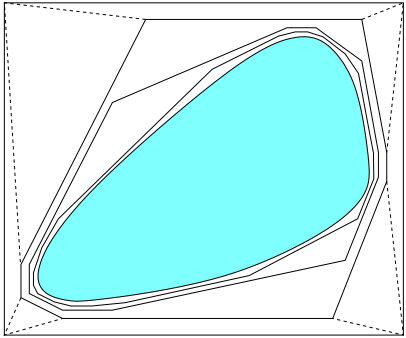,height=1.75in}}
\caption{An inflated hierarchy.  Only the outermost level of (dashed)
connecting edges is shown.  (Compare with Figure 1.)}
\label{F:inflated}
\end{figure}

As in the previous section, we first consider the special case of a
single point $p$ moving around a polygon~$Q$.  Let $Q_0, Q_1, \dots$
be the Dudley hierarchy of $Q$, and let $\e_i = \alpha D/2^i$, the
approximation error of $Q_i$ given by Lemma~\ref{L:compass}(a).  (We
can obtain similar results using a Dudley hierarchy instead.)  We
define an \emph{inflated} hierarchies $Q'_0, Q'_1, \dots$, where $Q'_i
= Q_i[\e_i]$.  See Figure~\ref{F:inflated} (but ignore the dashed edges
for now).  Since $Q_0$ is a rectangle, no envelope $Q_i$ has an acute
vertex angle, so $d(Q'_i,Q) \le (1+\sqrt{2})\e_i$.

As the point moves, we maintain a separating edge $e$ of some
uninflated envelope $Q_i$.  The separation certificate expires when
$p$ crosses the line $\Line{e}$.  To compute the new separation
certificate, we find the index $j$ such that $p$ lies outside $Q'_j$
but inside $Q'_{j-1}$.  For the new separating edge, we take the edge
of $Q_j$ parallel to the edge of $Q'_j$ that intersects $\Line{e}$.
To keep the required space down to $O(n)$, we cannot precompute the
breakpoints; instead, we compute each breakpoint $\Line{e}\cap Q'_j$
on the fly in $O(j)$ time when we need it.  The total cost of an event
is $O(\log\mu)$, and the total number of events is the same as for the
uninflated compass hierarchy.

Alternately, we can connect successive levels in the inflated
hierarchy, as shown by the dashed edges in Figure \ref{F:inflated},
decomposing $Q_0'\setminus Q$ into a complex of triangles and
trapezoids.  Each cell in this complex has one inner edge and two or
three outer edges. After any event, we locate the cell $c$ in the
inflated complex containing $p$; if the inner edge of $c$ is an edge
of $Q'_j$, we take the corresponding edge of $Q_j$ to be the new
separating edge.  The update time and number of events is the same as
above.

\begin{lemma}
\label{L:hysteresis}
After any event, if $d(P,Q) = \Omega(D/n)$, then $P$ must move at
least $d(P,Q)/\beta$ before the next event, where $\beta =
2(1+\sqrt{2}) \approx 4.8284$.
\end{lemma}

%

The only time we do not obtain hysteresis is when the new separating
edge is an edge of the actual polygon $Q$, or equivalently, when the
point $p$ lies inside the polygonal annulus $Q[\e]\setminus Q$ for
some $\e = O(D/n)$.

For any real $\kappa>1$, say that a circle $C$ is
\emph{$\kappa$-clear} if its radius is at most $1/\kappa$ times the
distance from its center to the polygon $Q$.  A \emph{$\kappa$-clear
decomposition} of a path $\pi$ is a decomposition of $\pi$ into
contiguous segments $\pi_1, \pi_2, \dots, \pi_k$, such that each
segment $\pi_i$ is contained in a $\kappa$-clear disk.  The size of
such a decomposition is the number of segments.

\begin{theorem}
\label{Th:clear}
If $p$ moves along a path $\pi$ whose minimum distance to $Q$ is
$\Omega(D/n)$, then the number of events is at most the size of the
smallest $\kappa$-clear decomposition of $\pi$, where $\kappa =
2\beta+1 = 5 + 4\sqrt{2} \approx 10.6569$.
\end{theorem}

The constants $\beta$ and $\kappa$ are function of the minimum
external angle of any envelope and the ratio $\e_i/\e_{i-1}$.  By
using more complicated bounding polygons as the outermost level of the
hierarchy, and by letting the inflation offset grow more slowly, we
can decrease $\beta$ arbitrarily close to $1$ and $\kappa$ arbitrarily
close to~$3$.  Somewhat paradoxically, however, these modifications
\emph{increase} both the update time per event and the worst-case
number of events.  In particular, using an inflated Dudley hierarchy
instead of an inflated compass hierarchy doubles the value of $\beta$,
even though it leads to asymptotically fewer events in the worst case.


For the case of two translating polygons $P$ and $Q$, we construct
separate inflated hierarchies for both $P$ and~$Q$ and use a technique
similar to the lazy structure in Section \ref{ss:translation}.  The
separation certificate consists of an vertex~$v$ of $P_i$ and an edge
$e$ of $Q_i$, or vice versa, for some~$i$.  When $v$ hits $e$, we
compute a new separation certificate based on which inflated envelope
$Q'_j$ contains the vertex~$v$.  The resulting structure exhibits
hysteresis and path-sensitivity (with different constants $\beta$
and~$\kappa$) and still satisfies Theorem~\ref{Th:transpoly}.

Finally, a further modification gives us \emph{rotational} hysteresis
as well.  That is, after any event, $P$ must either move a distance of
$\Omega(d(P,Q))$ or rotate by an angle of $\Omega(d(P,Q))$ relative to
$Q$ before the next event occurs.  This modification requires a
\emph{lower} bound on the external angles of any envelope, so we can
use only the compass hierarchy, not the Dudley hierarchy.  Rotational
hysteresis implies that the number of events is bounded by the size of
a $\kappa$-clear decomposition of the path that the polygons traverse through
the three-dimensional configuration space, for some constant $\kappa$.

We omit further details from this version of the paper.

\section{Conclusions and Open Problems}
\label{S:outro}

As we mentioned in the introduction, we do no foresee any major
obstacles to generalizing the two-dimensional results in this paper to
three dimensions.  We already have a few preliminary results, which we
plan to develop further in the full paper.  Generalizations of both
the compass hierarchy and the Dudley hierarchy are easy to define, and
at least the Dudley hierarchy provides good approximation bounds.
After constructing the Dudley hierarchy of two polyhedra, we can
maintain a separation certificate in $O(\polylog(D/s))$ time per
event, using an approach similar to the lazy structure in Section
\ref{S:sensitive}.  The number of events is $O(\mu^2)$ for algebraic
translation, and $O(\mu^4)$ for algebraic rigid motion.

Returning to our two-dimensional results, for various measures of
kinetic efficiency, we have KDSs with good performance under that
measure, but only at the expense of some other kinetic quality.  It
would be desirable to find a single KDS that uses only linear space
per polygon, has good event bounds under all three classes of motion
(convex translation, algebraic translation, and algebraic rigid
motion), and exhibits both translational and rotational hysteresis.

Finally, we intend to work on extending the two body methods (narrow
phase) of this paper to multiple moving convex polytopes. This will
require a kinetic structure that implements the broad phase of
collision detection and determines which pairs of objects need to be
passed on to the narrow phase.

\bigskip\noindent
\textbf{Acknowledgments:}
We wish to thank Julien Basch for fruitful discussions, and Pankaj
Agarwal for making us aware of Dudley's result \cite{Dud}.

\def\burl#1{$\langle$\url{#1}$\rangle$}

\end{document}